\documentstyle[12pt,graphicx,preprint,aps,tighten]{revtex}
\begin{document}
\draft

\preprint{
\parbox{4cm}{\vspace*{2cm}}}

\title{Gravitino condensation in fivebrane backgrounds}
\author{Noriaki Kitazawa
\thanks{e-mail: kitazawa@phys.metro-u.ac.jp}}
\address{Department of Physics, Tokyo Metropolitan University,
         Hachioji, Tokyo 192-0397, Japan}
\date{\today}
\maketitle

\begin{abstract}

We calculate
 the tension of the D3-brane in the fivebrane background
 which is described by the exactly solvable
 SU$(2)_k \times$U$(1)$ world-sheet conformal field theory
 with large Ka\v c-Moody level $k$.
The D3-brane tension is extracted
 from the amplitude of one closed string exchange
 between two parallel D3-branes,
 and the amplitude is calculated
 by utilizing the open-closed string duality.
The tension of the D3-brane in the background
 does not coincide with the one in the flat space-time
 even in the flat space-time limit: $k \rightarrow \infty$.
The finite curvature effect
 should vanish in the flat space-time limit
 and only the topological effect can remain.
Therefore,
 the deviation suggests
 the condensation of gravitino and/or dilatino
 which has been expected in the fivebrane background
 as a gravitational instanton.

\end{abstract}

\newpage

\section{Introduction}
\label{sec:intro}

It has been pointed out that
 the local supersymmetry can be spontaneously broken
 by the gravitino condensation due to the topological effect
 of the quantum gravity\cite{Witten,KMP}.
An explicit calculation
 in the framework of the Euclidean quantum gravity theory
 is given in Ref.\cite{KMP},
 and the finite value of the gravitino pair condensate
 $\langle \psi_{\mu\nu} \psi_{\mu\nu} \rangle$ is obtained,
 where $\psi_{\mu\nu}=D_\mu\psi_\nu-D_\nu\psi_\mu$
 is the field strength of the gravitino field $\psi_\mu$
 and $D_\mu$ is the covariant derivative.
This mechanism is very attractive, 
 since it is less model-dependent.

It is claimed that
 this mechanism might work
 also in the low energy effective theory of the superstring theory.
The existence of the appropriate zero modes
 for the gravitino pair condensation
 in the fivebrane background is explicitly shown in Ref.\cite{Rey}
 in the framework of the Euclidean quantum gravity theory.
Since the fivebrane background is exactly described
 by using the world-sheet conformal field theory\cite{CHS,AFK,Kounnas} 
 as a solution of the string theory,
 it is interesting to confirm this suggestion
 in the string theory from the first principle. 

A D-brane act as a source for closed strings.
The D-brane tension
 is nothing but the coupling strength
 between a D-brane and a single lowest-lying closed string state.
In the low-energy limit
 (considering only the lowest-lying (massless) states of
 the string excitation)
 a single Dp-brane in the type II string theory
 is described by the effective action
\begin{equation}
 S_p = - T_p \int d^{p+1} \xi
  \left\{
   e^{-\Phi}
    \left[
     - \det \left( G_{ab} + B_{ab} + 2 \pi \alpha' F_{ab} \right)
    \right]^{1/2}
    + {\cal O}(\alpha') 
  \right\},
\label{Dp-action}
\end{equation}
 where $T_p$ is the tension of the Dp-brane,
 $\xi^a$ is the Dp-brane coordinate ($a=0,1,\cdots,p$),
 $\Phi$ is the dilaton field,
 $G_{ab}$ is the pull back of the string metric $G_{\mu\nu}$
 ($\mu,\nu = 0,1,\cdots,9$),
 $B_{ab}$ is the pull back of the B-field $B_{\mu\nu}$,
 and $F_{ab}$ is the gauge field from the open string in the Dp-brane.
(Here, we neglect the contribution from R-R fields.)
After rescaling the string metric to the Einstein metric
 $g_{\mu\nu}=e^{-\phi/2} G_{\mu\nu}$
 with $\Phi=\langle \Phi \rangle + \phi$,
 the couplings of a single dilaton or graviton to a Dp-brane
 can be obtained.
The coupling is proportional to the effective Dp-brane tension
 $\tau_p = T_p e^{-\langle \Phi \rangle}$.

\begin{figure} \centering
\includegraphics[width=0.8\textwidth]{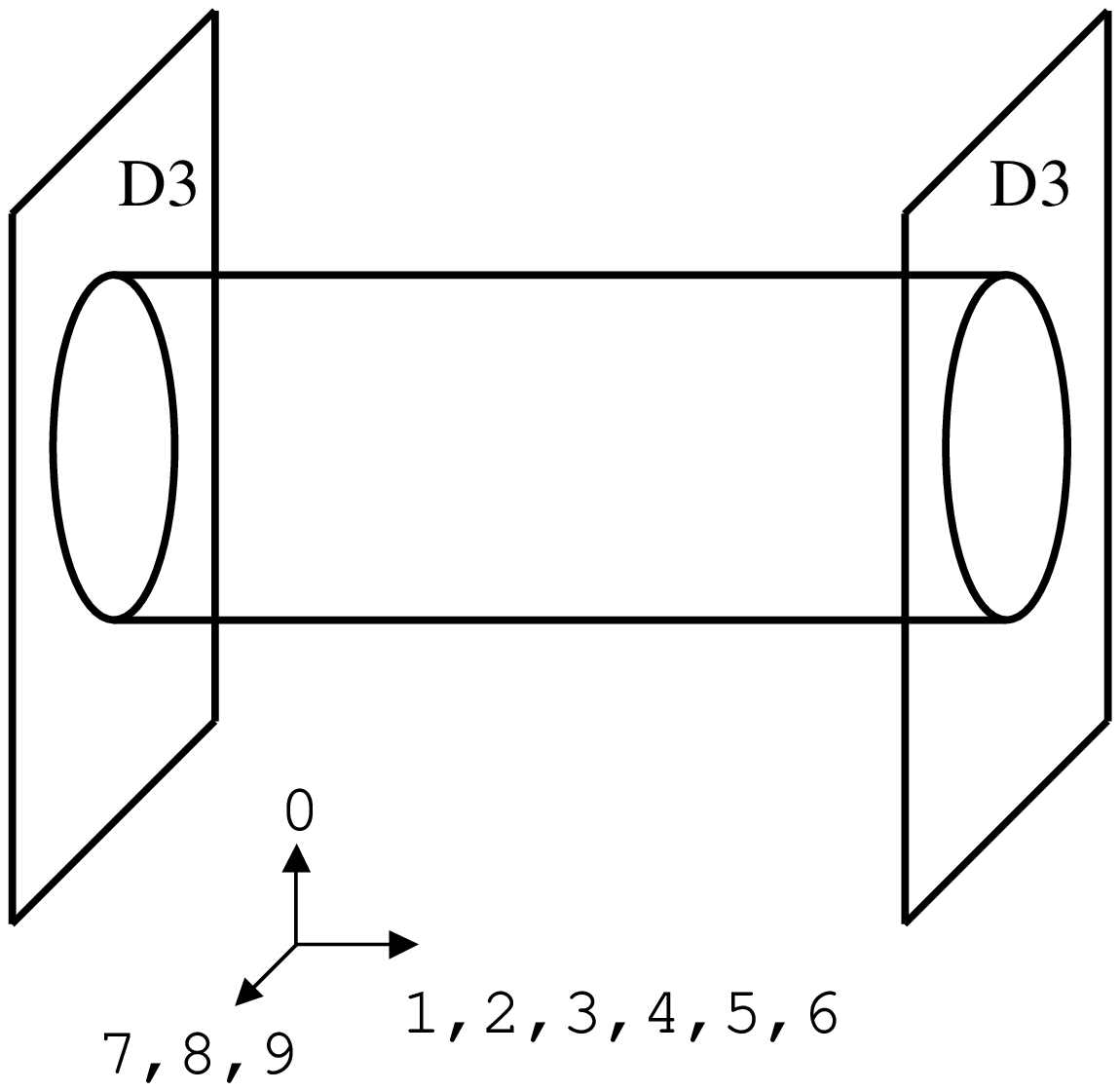}
\caption{
 The diagram of 
  one closed string exchange between two parallel D3-branes}
\label{fig:diagram}
\end{figure}

In case of the curved space-time background
 the higher order terms in Eq.(\ref{Dp-action})
 contribute to the effective tension.
The higher order $R^2$ terms, for example,
 are obtained in Ref.\cite{BBG} in case of D3-brane:
\begin{equation}
 S_3^{R^2}
 = - T_3 {{(2 \pi \alpha')^2} \over {96}}
   \int d^4 \xi e^{-\Phi} \mbox{tr} \left( R \wedge {}^*R \right)
 + \cdots,
\label{higher1}
\end{equation}
 where
 $R^c{}_d = {1 \over 2} R^c{}_{dab} d\xi^a \wedge d\xi^b$
 and
 ${}^*R^c{}_d = {1 \over 4} \sqrt{-G}
  \epsilon_{ab}{}^{a'b'} R^c{}_{da'b'} d\xi^a \wedge d\xi^b$.
The non-trivial background
 gives rise to a vacuum expectation value of the curvature tensor,
 and the effective D-brane tension is modified.
The supersymmetric counter part of
 the effective action of Eq.(\ref{higher1}) can be obtained
 by using the superfield method described in Ref.\cite{BT,CGNW,APS,HW}.
We find
\begin{equation}
 S_3^{\psi^2}
 \propto
  T_3 {{(2 \pi \alpha')^2} \over {96}}
  \int d^4 \xi \sqrt{-G} e^{-\Phi}
  \Big( \partial^c \lambda \partial_c \lambda \Big)
  \Big( \psi^{ab} \psi_{ab} \Big)
 + \cdots,
\label{higher2}
\end{equation}
 where $\psi$ is a linear combination of two gravitinos
 corresponding to $N=2$ supersymmetry of the type II theory,
 and $\lambda$, which has mass dimension $-1/2$,
 is a Goldstone fermion
 associated with the half supersymmetry breaking by the D3-brane.
The vacuum expectation value of composite operators, like
 $\langle \big( \partial^c \lambda \partial_c \lambda \big)
          \big( \psi^{ab} \psi_{ab} \big) \rangle$
 in this example,
 also modify the effective D-brane tension.
Since the low-energy amplitude of
 one closed string exchange between two parallel D-branes
 (Fig.\ref{fig:diagram})
 is determined by the effective D-brane tension,
 the non-trivial background effect may modify
 the low-energy amplitude.

In this paper
 we calculate the amplitude of Fig.\ref{fig:diagram}
 in the fivebrane background,
 and compare it with the same amplitude in the flat space-time. 
Since the superstring theory
 in the fivebrane background, $M^6 \otimes W^{(4)}_k$,
 where $M^6$ is the six-dimensional flat Minkowski space-time
 and $W^{(4)}_k \equiv$ SU$(2)_k \times $U$(1)$ is the four-dimensional
 non-trivial space,
 can be exactly constructed,
 we can calculate the amplitude exactly.
In the flat space-time limit, $k \rightarrow \infty$,
 the finite curvature effect disappears in the amplitude,
 and only the topological effect should remain.
(Notice also that the space-time background interpretation
 of this string solution exists only for large Ka\v c-Moody level $k$.)
We will see in the low-energy limit that
 the amplitude in the flat space-time limit
 does not coincide with the one in the flat space-time.
Namely,
 the effective D-brane tension in the flat space-time limit
 does not coincide with the one in the flat space-time.
This suggests the gravitino and/or dilatino condensation
 by the topological effect in the string theory.

In the next section
 we review the string solution in the fivebrane background.
In section \ref{sec:amplitude}
 the amplitude of Fig.\ref{fig:diagram} is calculated
 both in the flat space-time and in the fivebrane background.
The flat space-time limit of the amplitude is discussed.
It will be found that
 the effective D-brane tension in the flat limit
 does not coincide with the one in the flat space-time.
In the last section we summarize the result.

\section{String Theory on Fivebrane Backgrounds}
\label{sec:fivebrane}

In this section
 we briefly review the exact string solution
 in the fivebrane background
 (see  Refs.\cite{CHS,AFK} for details).
We concentrate on the construction of
 the partition functions of closed and open strings
 in the fivebrane background $M^6 \otimes W^{(4)}_k$.

In case of the flat space-time the world-sheet theory consists of
 a free bosonic field $X^\mu(z,{\bar z})$
 and two free fermionic fields
 $\psi^\mu(z)$ and ${\tilde \psi}^\mu({\bar z})$.
The system has $(N,{\tilde N})=(1,1)$ superconformal symmetry
 with the central charge $c_m={\tilde c}_m=10+10/2=15$
 which is cancelled by the ghost contribution
 $c_g={\tilde c}_g=-26+11=-15$.
The non-trivial background $M^6 \otimes W^{(4)}_k$
 can be described by replacing unconstrained fields
 $X^\mu$ with $\mu=6,7,8,9$
 to the fields constrained on the group manifold
 $W^{(4)}_k =$ SU$(2)_k \times$U$(1)$.
Namely,
 a part of the world-sheet theory
 corresponding to the space-time dimensions of $\mu=6,7,8,9$
 is replaced by the SU$(2)_k \times$U$(1)$ WZW model
 with $(N,{\tilde N})=(4,4)$ superconformal symmetry
 which has the same central charge of the replaced part
 $c_{WZW}=6$.

The holomorphic part of the SU$(2)_k \times$U$(1)$ WZW model
 is described by three SU$(2)_k$ bosonic currents $J_i(z)$
 ($i = 1,2,3$), one free bosonic current $J_4(z)=\partial X^{\mu=6}$
 and four free fermionic fields $\Psi_a$ ($a=1,2,3,4$).
These currents and fields satisfy
 the following the operator product expansion.
\begin{eqnarray}
 J_i(z) J_j(z') &\sim&
  - { k \over 2} {{\delta_{ij}} \over {(z-z')^2}}
  + \epsilon_{ijl} {{J_l} \over {z-z'}},
\\
 J_4(z) J_4(z') &\sim& - {1 \over {(z-z')^2}},
\\
 \Psi_a(z) \Psi_b(z') &\sim& - {{\delta_{ab}} \over {z-z'}}, 
\end{eqnarray}
 where $k$ is the Ka\v c-Moody level of SU$(2)_k$.
The $N=4$ superconformal symmetry transformation
 is generated by the following energy-momentum tensor $T(z)$,
 supercurrents $G_a(z)$ and SU$(2)_n$ currents $S_i(z)$.
\begin{eqnarray}
 T &=&
  - {1 \over 2}
  \left(
   {2 \over {k+2}} J_i^2 + J_4^2
   - \Psi_a \partial \Psi_a + Q \partial J_4
  \right),
\\
 G_i &=&
  \sqrt{{2 \over {k+2}}}
   \left(
    J_i \Psi_4 - \epsilon_{ijl} J_j \Psi_l
    + {1 \over 2} \epsilon_{ijl} \Psi_4 \Psi_j \Psi_l
   \right)
  - J_4 \Psi_i - Q \partial \Psi_i,
\\
 G_4 &=&
  \sqrt{{2 \over {k+2}}}
   \left(
    J_i \Psi_i + {1 \over {3!}} \epsilon_{ijl} \Psi_i \Psi_j \Psi_l
   \right)
  + J_4 \Psi_4 + Q \partial \Psi_4,
\\
 S_i &=&
  {1 \over 2}
  \left(
   \Psi_4 \Psi_i + {1 \over 2} \epsilon_{ijl} \Psi_j \Psi_l
  \right).
\end{eqnarray}
The background charge $Q$
 determines the gradient of the linear-dilaton background
 $\Phi=QX^{\mu=6}$.
The world-sheet field $X^{\mu=6}(z,{\bar z})$
 is called Feigin-Fuchs field.
The value $Q=\sqrt{2/(k+2)}$ is required
 for the correct central charge of $c_{WZW}=6$,
 and the Ka\v c-Moody level $n$ of SU$(2)_n$ is fixed to unity.
(In the present discussion we set $\alpha'=2$.
For example, $Q=\sqrt{\alpha'/(k+2)}=\sqrt{2/(k+2)}$.)
The anti-holomorphic part has exactly the same structure.

\subsection{The closed string partition function}
\label{sebsec:closed}

The partition function of the closed string is defined by
\begin{equation}
 Z^c(\tau)
 =
  \mbox{\rm Tr}
  \left[
   e^{2 \pi i \tau_1 P - 2 \pi \tau_2 H}
  \right]
 =
  \left( q {\bar q} \right)^{-d/24}
  \mbox{\rm Tr}
  \left[ q^{L_0} {\bar q}^{{\tilde L}_0} \right],
\end{equation}
 where $P=L_0-{\tilde L}_0$
 and $H=L_0+{\tilde L}_0-d/24$ are
 the world-sheet momentum and Hamiltonian operators,
 $\tau = \tau_1 + i \tau_2$, $q = e^{2 \pi i \tau}$,
 $d = (c + {\tilde c})/2$ with total central charges $c$ and ${\tilde c}$
 and $L_0$ and ${\tilde L}_0$ are Virasoro generators.
The torus vacuum amplitude is obtained
 by integrating this partition function
 over the moduli parameter $\tau$
\begin{equation}
 A_{T^2} = \int_F {{d^2\tau} \over {4\tau_2}} Z^c(\tau),
\end{equation}
 where $F$ denotes the fundamental region of the moduli space.
The partition function
 can be decomposed in two parts $Z^c = Z^c_B \times Z^c_F$,
 where $Z^c_B$ is the bosonic
 ($X^\mu$ and $bc$-ghosts) contribution
 and $Z^c_F$ is the fermionic
 ($\psi^\mu$, ${\tilde \psi}^\mu$ and $\beta\gamma$-ghosts) contribution.

The construction of the partition function
 in fivebrane backgrounds can be understood
 by starting with the partition function in the flat space-time.
In case of $D=10$ dimensional flat space-time
 the bosonic contribution $Z^c_B$ is obtained as
\begin{equation}
 Z^c_B(\tau) =
 {{i V_D} \over {4\pi^2 \alpha' \tau_2}}
 \left( Z^c_X(\tau) \right)^{D-2},
\label{bosonic}
\end{equation}
 where
\begin{equation}
 Z^c_X(\tau) = (4\pi^2 \alpha' \tau_2)^{-1/2} |\eta(\tau)|^{-2}
\label{single-bosonic}
\end{equation}
 is the contribution of one space-time dimension.
The factor $i$ in Eq.(\ref{bosonic}) originates from the negative sign
 of the time component of the flat space-time Minkowski metric.
The fermionic contribution is obtained as
\begin{equation}
 Z^c_F(\tau) = Z_\psi(\tau) \left( Z_\psi(\tau) \right)^*,
\label{fermionic}
\end{equation}
 where
\begin{equation}
 Z_\psi(\tau) =
 {1 \over 2}
 \sum_{\alpha,\beta}
 (-1)^{\alpha+\beta-\alpha\beta}
 \left(
 {\theta{ \alpha \choose \beta}(\tau)
  \over
  {\eta(\tau)}}
 \right)^{(D-2)/2}
\label{fermionic-holomorphic}
\end{equation}
 for the type IIB theory.
The two sectors, Neveu-Schwarz and Ramond sectors,
 are included by the summation over $\alpha = 0,1$.
The GSO projection is realized
 by the summation over $\beta = 0,1$.

We modify the above result
 to the case of the fivebrane background $M^6 \otimes W^{(4)}_k$.
The bosonic contribution of the $M^6$ part can be obtained
 by setting $D=6$ in Eq.(\ref{bosonic}).
The contribution of the Feigin-Fuchs field
 (U$(1)$ part of $W^{(4)}_k$)
 is given by Eq.(\ref{single-bosonic}) times $V_1$
 as the contribution of a single space-like coordinate,
 if the state of momentum $k$ is defined as
\begin{equation}
 | k \rangle =
 : e^{i (k+i{Q \over {\alpha'}}) X^{\mu=6}(0,0)} : | 0 \rangle
\end{equation}
 corresponding to the modification of $L_0$ and ${\tilde L}_0$.
By the effect of the Feigin-Fuchs field,
 the whole mass spectrum
 is uniformly lifted up by $\Delta m^2=(Q/\alpha')^2$,
 and there is no massless state.

The bosonic contribution of the SU$(2)_k$ part
 should be described by the Virasoro character
\begin{equation}
 \chi_k^n(\tau)
 =  q^{s_{n,k}} \mbox{\rm ch}_{n,k} (\tau),
\end{equation}
 where $n=0,1,\cdots$ denotes the representation of SU$(2)$,
 $s_{n,k}={{((n+1)/2)^2} \over {k+2}} - {3 \over {24}}$
 and $\mbox{\rm ch}_{n,k} (\tau)$ is the SU$(2)_k$ character
 which corresponds to $\mbox{\rm Tr} (q^{L_0})$
 (see Refs.\cite{KP,GW} for precise definitions).
The actual form of the Virasoro character is
\begin{equation}
 \chi_k^n(\tau) =
  {1 \over {(\eta(\tau))^3}} 
  \cdot 2(k+2)
  \sum_{m=-\infty}^\infty
  \left( m + {{n+1} \over {2(k+2)}} \right)
  e^{2 \pi i (k+2) \tau \left( m + {{n+1} \over {2(k+2)}} \right)^2}.
\label{character}
\end{equation}
The Virasoro character with $n \leq k$
 is a unitary representation under the modular transformation:
\begin{eqnarray}
 \chi_k^n(-1/\tau) &=& \sum_{n'=0}^k S_{nn'} \chi_k^{n'}(\tau),
\qquad
 S_{nn'} = \left( {2 \over {k+2}} \right)^{1/2}
           \sin \left( \pi {{(n+1)(n'+1)} \over {k+2}} \right),
\label{modular-s}\\
 \chi_k^n(\tau+1) &=& \sum_{n'=0}^k T_{nn'} \chi_k^{n'}(\tau),
\qquad
 T_{nn'} = \delta_{nn'}
             e^{i {\pi \over 2} 
                \left( {{(n+1)^2} \over {k+2}}
                       - {1 \over 2} \right)}.
\label{modular-t}
\end{eqnarray}
Therefore, a combination
\begin{equation}
 \sum_{n=0}^k \chi_k^n(\tau) \left( \chi_k^n(\tau) \right)^*
\end{equation}
 is a modular invariant,
 but this is not appropriate for the SU$(2)_k$ contribution
 to the partition function,
 since the whole partition function
 (or the vacuum amplitude) should be modular invariant.

The modular invariance and 
 the extraction of the physical state by the GSO projection
 lead the following partition function for even $k$.
\begin{eqnarray}
 Z^c(\tau) &=&
  {{i V_6} \over {4\pi^2 \alpha' \tau_2}}
  \left( Z^c_X(\tau) \right)^4
 \cdot
  V_1 Z^c_X(\tau)
\nonumber\\
  && \times
  \sum_{\alpha, \beta} \sum_{{\bar \alpha}, {\bar \beta}}
  \sum_{\gamma, \delta}
  {1 \over 2}
  (-1)^{\alpha+\beta-\alpha\beta}
  \left(
  {\theta{\alpha \choose \beta}(\tau)
   \over
   {\eta(\tau)}}
  \right)^2
  \left(
  {\theta{{\alpha+\gamma} \choose {\beta+\delta}}(\tau)
   \over
   {\eta(\tau)}}
  \right)^2
\nonumber\\
 && \times
  {1 \over 2}
  (-1)^{{\bar \alpha}+{\bar \beta}-{\bar \alpha}{\bar \beta}}
  \left(
  \left(
  {\theta{{{\bar \alpha}} \choose {{\bar \beta}}}(\tau)
   \over
   {\eta(\tau)}}
  \right)^2
  \left(
  {\theta{{{\bar \alpha}+\gamma} \choose {{\bar \beta}+\delta}}(\tau)
   \over
   {\eta(\tau)}}
  \right)^2
  \right)^* 
\nonumber\\
 && \times
  {1 \over 2}
  (-1)^{\delta (\alpha+{\bar \alpha} + \gamma k / 2)}
  Z_k
   \Big[
    \begin{array}{c}
     \gamma \\
     \delta
    \end{array}
   \Big](\tau),
\label{partition-closed}
\end{eqnarray}
 where
\begin{equation}
 Z_k
  \Big[
   \begin{array}{c}
    \gamma \\
    \delta
   \end{array}
  \Big](\tau)
 = \sum_{n=0}^k
   e^{i \pi \delta n}
   \chi_k^n (\tau)
   \left( \chi_k^{n+\gamma(k-2n)} (\tau) \right)^*
\end{equation}
 which covariantly transforms under the modular transformation as
\begin{eqnarray}
 Z_k
  \Big[
   \begin{array}{c}
    \gamma \\
    \delta
   \end{array}
  \Big](-1/\tau)
 &=&
 e^{i \pi \gamma \delta k}
 Z_k
  \Big[
   \begin{array}{c}
    \delta \\
    \gamma
   \end{array}
  \Big](\tau),
\\
 Z_k
  \Big[
   \begin{array}{c}
    \gamma \\
    \delta
   \end{array}
  \Big](\tau+1)
 &=&
 e^{-i \pi \gamma^2 k/2}
 Z_k
  \Big[
   \begin{array}{c}
    \gamma \\
    \delta+\gamma
   \end{array}
  \Big](\tau).
\end{eqnarray}
In Eq.(\ref{partition-closed})
 the summation over $\delta = 0, 1$
 realizes the additional GSO projection
 which reduces the space-time supersymmetry by a factor of $2$.
This corresponds to the fact that
 the fivebrane configuration is a BPS saturated state
 which preserves only half of the supersymmetry of the theory.
The fermionic contribution
 is the same in case of the flat space-time,
 except for the modification of the contribution
 from the fermions in $W^{(4)}_k$ by the additional GSO projection.
The summation over $\gamma = 0, 1$
 is required by the modular invariance,
 and a twisted sector ($\gamma=1$) should exist.

We discuss the flat space-time limit $k \rightarrow \infty$
 of the partition function of Eq.(\ref{partition-closed}).
Consider first
 the $k \rightarrow \infty$ limit of the Virasoro character
 of Eq.(\ref{character}).
Since $\tau_2 \geq 1$ in the fundamental region of the moduli space,
 there is an exponential suppression factor
 in the summation over $m$.
Considering $(n+1)/2(k+2)<1/2$,
 a contribution with $m=0$ dominates in $k \rightarrow \infty$ limit.
Therefore,
\begin{equation}
 \lim_{k \rightarrow \infty} \chi_k^n(\tau) =
  {1 \over {(\eta(\tau))^3}} 
  \cdot (n+1)
  e^{2 \pi i (k+2) \tau \left({{n+1} \over {2(k+2)}} \right)^2}.
\label{limit-character}
\end{equation}
Then,
\begin{equation}
 \lim_{k \rightarrow \infty}
 Z_k
  \Big[
   \begin{array}{c}
    0 \\
    0
   \end{array}
  \Big](\tau)
 = {1 \over {|\eta(\tau)|^{-6}}}
   \lim_{k \rightarrow \infty} \sum_{n=0}^k
   2(k+2)
   \left( {{n+1} \over \sqrt{2(k+2)}} \right)^2
   e^{-2 \pi \tau_2 \left( {{n+1} \over \sqrt{2(k+2)}} \right)^2}.
\end{equation}
We can replace the summation to the integration
 over the valuable $x=(n+1)/\sqrt{2(k+2)}$:
\begin{eqnarray}
 \lim_{k \rightarrow \infty}
 Z_k
  \Big[
   \begin{array}{c}
    0 \\
    0
   \end{array}
  \Big](\tau)
 &=& {1 \over {|\eta(\tau)|^{-6}}}
     \lim_{k \rightarrow \infty}
     2(k+2)
     \int_0^{{k \over {\sqrt{2(k+2)}}}}
     {{dx} \over {1/\sqrt{2(k+2)}}}
     x^2 e^{- 2 \pi \tau_2 x^2}
\nonumber\\
 &=& {1 \over {|\eta(\tau)|^{-6}}}
     \lim_{k \rightarrow \infty}
     \left( 2(k+2) \right)^{3/2}
     \int_0^\infty
     dx x^2 e^{- 2 \pi \tau_2 x^2}
\nonumber\\
 &=& \left( Z_X(\tau) \right)^3
     \lim_{k \rightarrow \infty} 2 \pi^2 \left( {2 \over Q} \right)^3,
\label{limit-Zk}
\end{eqnarray}
 where we use $\alpha'=2$ in the closed string Virasoro character
 due to the relation
 between mode operators and the space-time momentum operator:
 $\alpha_0^\mu = {\tilde \alpha}_0^\mu = \sqrt{\alpha' / 2} p^\mu$.
The factor $2 \pi^2 (2/Q)^3$
 is the volume of $S^3$ of the radius $2/Q$,
 which is the volume of the SU$(2)_k$ manifold.
Since $Q \rightarrow 0$ in $k \rightarrow \infty$ limit,
 the factor becomes $V_3$ in the flat space-time limit,
 and Eq.(\ref{limit-Zk}) becomes just
 the contribution of three space-like coordinates
 in the flat space-time.
We can show in the same way that
 $Z_k {\gamma \brack \delta} (\tau)$
 with other values of $\gamma$ and $\delta$
 vanishes in $k \rightarrow \infty$ limit.
Therefore,
 we see that the partition function of Eq.(\ref{partition-closed})
 reduces to the one in the flat space-time
 with an additional $1/2$ factor
 in the flat space-time limit $k \rightarrow \infty$.
This additional factor
 is the remnant of the additional GSO projection.
The supersymmetry
 which is broken by the fivebrane background
 does not recover in the simple flat space-time limit.
This suggests that the supersymmetry
 is broken by some topological effects.

\subsection{The open string partition function}
\label{sebsec:open}

The argument which is similar to the closed string
 is also possible for the open string.
The partition function of the (oriented) open string is defined by
\begin{equation}
 Z^o(t)
 = \mbox{\rm Tr} \left[ e^{- 2 \pi t H} \right]
 = q^{-c/24} \mbox{\rm Tr} \left[ q^{L_0} \right],
\end{equation}
 where $H = L_0 - c/24$ is the world-sheet Hamiltonian,
 $q = e^{2 \pi i \tau}$ with $\tau = it$
 and we use the doubling trick.
The cylinder vacuum amplitude
 is obtained by integrating this partition function over $t$.
\begin{equation}
 A_{C^2} = \int_0^\infty {{dt} \over {2t}} Z^o(t).
\end{equation}
The partition function can be separated in two parts:
 $Z^o(t) = Z^o_B(t) \times Z^o_F(t)$.
In case of the $D=10$ dimensional flat space-time
 the bosonic ($X^\mu$ and $bc$-ghosts) contribution is
\begin{equation}
 Z^o_B(t) = {{i V_D} \over {8 \pi^2 \alpha' t}}
            \left( Z_X(t) \right)^{D-2}
\label{bosonic-open}
\end{equation}
 with
\begin{equation}
 Z^o_X(t) = (8 \pi^2 \alpha' t)^{-1/2} (\eta(it))^{-1}.
\label{sigle-bosonic-open}
\end{equation}
The fermionic
 ($\psi^\mu$, ${\tilde \psi}^\mu$ and $\beta\gamma$-ghosts)
 contribution is
\begin{equation}
 Z^o_F(t) = Z_\psi(it),
\label{fermionic-open}
\end{equation}
 where $Z_\psi(\tau)$ is given in Eq.(\ref{fermionic-holomorphic}).
The Chan-Paton degrees of freedom
 can be included by multiplying a factor $n_{CP}^2$.

We modify the partition function in the flat space-time
 and obtain the one in the fivebrane background
 $M^6 \otimes W^{(4)}_k$.
The bosonic contribution of the $M^6$ part
 is Eq.(\ref{bosonic-open}) with $D=6$.
The contribution of the Feigin-Fuchs field
 is Eq.(\ref{sigle-bosonic-open}) times $V_1$.
The bosonic contribution of the SU$(2)_k$ part should be described by
 a linear combination of Virasoro characters $\chi_k^n$
 without their complex conjugate\cite{Cardy}.
The partition function in the fivebrane background
 can be obtained as
\begin{eqnarray}
 Z^o(t) &=&
 {{i V_6} \over {8 \pi^2 \alpha' t}}
 \left( Z^o_X(t) \right)^4
 \cdot
 V_1 Z^o_X(t)
\nonumber\\
 &&\times
 {1 \over 2}
 \sum_{\delta}
 \sum_{n=0}^k (n+1) (-1)^{\delta n} \chi_k^n(it)
\nonumber\\
 &&\times
 {1 \over 2}
 \sum_{\alpha,\beta}
 (-1)^{\alpha+\beta-\alpha\beta}
 \left(
 {\theta{ \alpha \choose \beta}(it)
  \over
  {\eta(it)}}
 \right)^2
 (-1)^{\delta\alpha}
 \left(
 {\theta{ \alpha \choose {\beta+\delta}}(it)
  \over
  {\eta(it)}}
 \right)^2,
\label{partition-open-fivebrane}
\end{eqnarray}
 where we consider the additional GSO projection
 in the same way in the closed string partition function.
The fermionic contribution
 is the same as the one in the flat space-time,
 except for the modification by the additional GSO projection.
The Chan-Paton degrees of freedom
 can be included by multiplying a factor $n_{CP}^2$.

In the flat space-time limit of $k \rightarrow \infty$
 the partition function reduces to the one in the flat space-time,
 except for an additional $1/2$ factor.
Consider the limit of the SU$(2)_k$ bosonic contribution
\begin{equation}
 Z_k^\delta(t) = \sum_{n=0}^k (n+1) (-1)^{\delta n} \chi_k^n(it)
\end{equation}
 using Eq.(\ref{limit-character}) in case of $\delta=0$:
\begin{eqnarray}
 \lim_{k \rightarrow \infty} Z_k^0(t)
 &=& \lim_{k \rightarrow \infty} \sum_{n=0}^k (n+1) \chi_k^n(it)
\nonumber\\
 &=& \lim_{k \rightarrow \infty} \sum_{n=0}^k (n+1)
     \cdot
     {1 \over {(\eta(it))^3}} 
     (n+1)
     e^{- 2 \pi (k+2) t \left({{n+1} \over {2(k+2)}} \right)^2}
\nonumber\\
 &=& {1 \over {(\eta(it))^3}}
     \lim_{k \rightarrow \infty} \sum_{n=0}^k
     2(k+2)
     \left( {{n+1} \over {\sqrt{2(k+2)}}} \right)^2
     e^{- \pi t \left({{n+1} \over {\sqrt{2(k+2)}}} \right)^2}.
\end{eqnarray}
The summation can be replaced by the integration
 over the valuable $x=(n+1)/\sqrt{2(k+2)}$.
\begin{eqnarray}
 \lim_{k \rightarrow \infty} Z_k^0(t)
 &=& {1 \over {(\eta(it))^3}}
     \lim_{k \rightarrow \infty}
     2(k+2)
     \int_0^{{k \over {\sqrt{2(k+2)}}}}
     {{dx} \over {1/\sqrt{2(k+2)}}}
     x^2 e^{- \pi t x^2}
\nonumber\\
 &=& {1 \over {(\eta(it))^3}}
     \lim_{k \rightarrow \infty}
     \left( 2(k+2) \right)^{3/2}
     \int_0^\infty dx x^2 e^{- \pi t x^2}
\nonumber\\
 &=& \left( Z_X(t) \right)^3 
     \lim_{k \rightarrow \infty} 2 \pi^2 \left( {2 \over Q} \right)^3,
\end{eqnarray}
 where we use $\alpha'=1/2$ in the open string Virasoro character
 due to the relation between a mode operator
 and the space-time momentum operator:
 $\alpha_0^\mu = \sqrt{2\alpha'} p^\mu$.
Therefore,
 $Z_k^0(t)$ reduces to just the contribution of
 three space-like coordinates in the flat space-time.
Since we can show in the same way that
 the flat limit of $Z_k^1(t)$ vanishes,
 the total partition function reduces to the one
 in the flat space-time with an $1/2$ factor.
This factor is the remnant of the additional GSO projection.
The meaning of this factor
 is the same of the additional factor
 in the closed string partition function.

\section{One Closed String Exchange Amplitude}
\label{sec:amplitude}

In this section
 we calculate the amplitude of Fig.\ref{fig:diagram}
 both in the flat space-time and in the fivebrane background.
The open string cylinder vacuum amplitude,
 which has been obtained in the previous section,
 can be utilized for the calculation
 under the open-closed string duality.

\subsection{The amplitude in the flat space-time}
\label{subsec:flat}

First,
 we calculate the amplitude of Fig.\ref{fig:diagram}
 in the flat space-time.
This calculation is described in detail in
Ref.\cite{Polchinski}. Consider two parallel D3-branes separated with
distance $y$
 and an (oriented) open string
 whose ends are fixed in different D3-branes.
Since the open string 
 has no space-time momentum in the directions of $\mu=1,2,3,4,5,6$
 and it is stretched between two D3-branes,
 in the world-sheet Hamiltonian
 a part of the momentum term
 $\alpha' p^\mu p_\mu$  with $\mu=1,2,3,4,5,6$
 is replaced by the tension term $y^2/4 \pi^2 \alpha'$.
Considering this replacement
 in the calculation of the cylinder vacuum amplitude $A_{c^2}$
 in the previous section
 gives the amplitude of Fig.\ref{fig:diagram}.
We should set the Chan-Paton factor $n_{CP}^2=2$
 to represent the left and right degrees of freedom 
 of the closed string.
The result is
\begin{equation}
 A^{M^{10}}(y) =
 \int_0^\infty {{dt} \over {2t}}
 {1 \over {V_6}}
 \left( 8 \pi^2 \alpha' t \right)^{6/2}
 e^{- {{y^2} \over {2 \pi \alpha'}} t}
 \cdot 2 Z^o_B(t)Z^o_F(t).
\end{equation}
The world-sheet time coordinate $t$ for the open string
 should be changed to the one for the closed string, $s=\pi/t$,
 by using the modular transformation:
\begin{eqnarray}
 \eta(it) &=& t^{-1/2} \eta(i/t),
\label{modular-eta}\\
 {\theta{\alpha \choose \beta}(it)
  \over
  {\eta(it)}}
 &=&
 {\theta{\beta \choose \alpha}(i/t)
  \over
  {\eta(i/t)}}.
\label{modular-theta}
\end{eqnarray}
Namely,
\begin{eqnarray}
 A^{M^{10}}(y) &=&
 {{i V_4} \over { 2 \pi \left( 8 \pi^2 \alpha' \right)^5}}
 \int_0^\infty ds
 \left( 8 \pi^2 \alpha' {\pi \over s} \right)^3
 e^{- {{y^2} \over {2 \pi \alpha'}} {\pi \over s}}
\nonumber\\
 && \times
 \left( \eta(is/\pi) \right)^{-8}
 \left\{
  Z_\psi^{NS-NS}(i s / \pi) + Z_\psi^{R-R}(i s / \pi)
 \right\},
\label{amplitude-flat}
\end{eqnarray}
 where
\begin{eqnarray}
 Z_\psi^{NS-NS}(i s / \pi)
 &=&
 {1 \over {\left( \eta(i s / \pi) \right)^4}}
 \left\{
 \left(
  \theta{0 \choose 0}(i s / \pi)
 \right)^4
 -
 \left(
  \theta{0 \choose 1}(i s / \pi)
 \right)^4
 \right\},
\label{NS-NS}\\
 Z_\psi^{R-R}(i s / \pi)
 &=&
 {1 \over {\left( \eta(i s / \pi) \right)^4}}
 \left\{
 -
 \left(
  \theta{1 \choose 0}(i s / \pi)
 \right)^4
 -
 \left(
  \theta{1 \choose 1}(i s / \pi)
 \right)^4
 \right\}
\label{R-R}
\end{eqnarray}
 represent the exchanges of the closed string in
 NS-NS and R-R modes, respectively.
The actual value of this amplitude is zero,
 which is required by the supersymmetry.
The contributions of the NS-NS and R-R exchanges are cancelled out.

The low-energy limit is the $s \rightarrow \infty$ limit
 in the integrant of the amplitude.
In this limit
 only the exchange of the closed string in lowest-lying states
 (graviton and dilaton in NS-NS sector)
 is extracted.
It is easy to show that
\begin{eqnarray}
 \lim_{s \rightarrow \infty}
  \left( \eta(i s / \pi) \right)^{-8}
  Z_\psi^{NS-NS}(i s / \pi) &=& 16 + {\cal O}(e^{-2s}),
\\
 \lim_{s \rightarrow \infty}
  \left( \eta(i s / \pi) \right)^{-8}
  Z_\psi^{R-R}(i s / \pi) &=& -16 + {\cal O}(e^{-2s}),
\end{eqnarray}
 where ${\cal O}(e^{-2s})$ is the contribution of massive states.
The low-energy limit
 of the NS-NS contribution of the amplitude is
\begin{eqnarray}
 A^{M^{10}}_{NS-NS}(y) &=&
 {{i V_4} \over { 2 \pi \left( 8 \pi^2 \alpha' \right)^5}}
 \int_0^\infty ds
 \left( 8 \pi^2 \alpha' {\pi \over s} \right)^3
 e^{- {{y^2} \over {2 \pi \alpha'}} {\pi \over s}}
 \cdot 16
\nonumber\\
 &=& i V_4 \cdot 2 \pi G_6(y),
\end{eqnarray}
 where
\begin{equation}
 G_d(y) 
  = {1 \over 4} \pi^{-{d \over 2}}
    \Gamma({{d-2} \over 2})
    y^{2-d} 
\end{equation}
 is the massless scalar Green's function in $d$ dimensions.
This amplitude corresponds to the diagram of
 the tree-level one graviton and one dilaton exchanges
 between D3-branes in the supergravity theory
 with the D-brane effective action of Eq.(\ref{Dp-action}).
The amplitude in the supergravity theory can be obtained as
\begin{equation}
 A^{M^{10}}_{NS-NS}(y)
 = i V_4 \cdot 2 \kappa^2 \tau_3^2 \cdot G_6(y),
\end{equation}
 where $\kappa$ is the gravitational constant in ten dimensions.
Therefore, 
 we have an expression for D3-brane tension
\begin{equation}
 \tau_3^2 = {\pi \over {\kappa^2}}.
\label{tension-flat}
\end{equation}
The same procedure is possible for general Dp-branes
 and we have
\begin{equation}
 \tau_p^2
 = {\pi \over {\kappa^2}} \left( 4 \pi^2 \alpha' \right)^{3-p}
\end{equation}
 which satisfies the T-duality relation 
\begin{equation}
 {{\tau_p} \over {\tau_{p-1}}}
 = {{1} \over {2 \pi \sqrt{\alpha'}}}.
\end{equation}

\subsection{The amplitude in fivebrane backgrounds}
\label{subsec:fivebrane}

The amplitude of Fig.\ref{fig:diagram}
 in the fivebrane background can be calculated
 in the same way in the previous subsection.

The transverse dimensions
 of two parallel D3-branes are $\mu=1,2,3,4,5,6$.
The non-trivial gravitational and B-field configurations
 are realized in the space of $\mu=7,8,9$,
 namely, the space in D3-brane world-volume.
The dilaton background has linear $X^{\mu=6}$ dependence,
 and the dilaton has a definite (but different)
 vacuum expectation value in each D3-brane world-volume.
The original supersymmetry of $\epsilon_L Q_L + \epsilon_R Q_R$,
 where $Q_L$ and $Q_R$ are left and right supercharges,
 is partially broken by this configuration.
This fivebrane background in type IIB theory
 preserves the supersymmetry of
\begin{equation}
 \epsilon_L
 = \Gamma^0 \Gamma^1 \Gamma^2 \Gamma^3 \Gamma^4 \Gamma^5
   \epsilon_L,
\qquad
\epsilon_R
 = - \Gamma^0 \Gamma^1 \Gamma^2 \Gamma^3 \Gamma^4 \Gamma^5
   \epsilon_R,
\end{equation}
 and D3-branes preserve the supersymmetry of
\begin{equation}
 \epsilon_L
 = \Gamma^0 \Gamma^7 \Gamma^8 \Gamma^9 \epsilon_R,
\end{equation}
 where $\Gamma^\mu$ is the gamma matrix in ten dimensions
 (see Ref.\cite{GK} for review).
Therefore,
 only eight ($1/4$) supersymmetries are preserved
 in this configuration.

Since the open string has no space-time momentum
 in the direction of $\mu=1,2,3,4,5,6$
 and it is stretched between two D3-branes,
 the terms $\alpha' p^\mu p_\mu$ with $\mu=1,2,3,4,5$
 and the term $\alpha' (p^{\mu=6} - i Q / \alpha')^2$ for $\mu=6$
 are replaced by a tension term $y^2 / 4 \pi^2 \alpha'$
 in the world-sheet Hamiltonian,
 where $y$ is the distance between two parallel D3-branes.
Namely,
 the amplitude is obtained by the replacement
\begin{equation}
 i V_7 \left( 8 \pi^2 \alpha' t \right)^{-7/2}
 \longrightarrow
 i V_1 \left( 8 \pi^2 \alpha' t \right)^{-1/2}
 q^{{{y^2} \over {4 \pi^2 \alpha'}}}
\end{equation}
 in Eq.(\ref{partition-open-fivebrane})
 with $n_{CP}^2=2$ and the integration over $t$:
\begin{eqnarray}
 A^{M^6 \otimes W_k^{(4)}}
 &=&
 \int_0^\infty {{dt} \over {2t}} \
 2 \cdot
 i V_1 \left( 8 \pi^2 \alpha' t \right)^{-1/2}
 e^{- {{y^2} \over {2 \pi \alpha'}} t}
 \left( \eta(it) \right)^{-5}
\nonumber\\
 &&\times
 {1 \over 2}
 \sum_{\delta}
 \sum_{n=0}^k (n+1) (-1)^{\delta n} \chi_k^n(it)
\nonumber\\
 &&\times
 {1 \over 2}
 \sum_{\alpha,\beta}
 (-1)^{\alpha+\beta-\alpha\beta}
 \left(
 {\theta{ \alpha \choose \beta}(it)
  \over
  {\eta(it)}}
 \right)^2
 (-1)^{\delta\alpha}
 \left(
 {\theta{ \alpha \choose {\beta+\delta}}(it)
  \over
  {\eta(it)}}
 \right)^2.
\end{eqnarray}
To understand this amplitude
 as the amplitude of a closed string exchange,
 the valuable $t$ should be changed to $s=\pi/t$.
By using Eqs. (\ref{modular-eta}), (\ref{modular-theta})
 and (\ref{modular-s}) we obtain
\begin{eqnarray}
 A^{M^6 \otimes W_k^{(4)}}
 &=&
 {{i V_4} \over { 2 \pi \left( 8 \pi^2 \alpha' \right)^5}}
 \int_0^\infty ds
 \left( 8 \pi^2 \alpha' {\pi \over s} \right)^3
 e^{- {{y^2} \over {2 \pi \alpha'}} {\pi \over s}}
\nonumber\\
 &&\times
 {{\left( 8 \pi^2 \alpha' \right)^{3/2}
   \left( \eta( is / \pi) \right)^3}
  \over
  {V_3}}
 \cdot
 {1 \over 2}
 \sum_{\delta}
 \sum_{n,n'} (n+1) (-1)^{\delta n} S_{nn'} \chi_k^{n'}(i s / \pi)
\nonumber\\
 &&\times
 \left( \eta(i s / \pi) \right)^{-8}
 \sum_{\alpha,\beta}
 (-1)^{\alpha+\beta-\alpha\beta}
 \left(
 {\theta{ \beta \choose \alpha}(i s / \pi)
  \over
  {\eta(i s / \pi)}}
 \right)^2
 (-1)^{\delta\alpha}
 \left(
 {\theta{ {\beta+\delta} \choose \alpha}(i s / \pi)
  \over
  {\eta(i s / \pi)}}
 \right)^2.
\end{eqnarray}
This amplitude
 should be compared with the amplitude of Eq.(\ref{amplitude-flat}).
The exchange of the twisted sector ($\delta=1$)
 is included in addition to the normal (untwisted) sector
 ($\delta=0$).

Now we take the flat space-time limit $k \rightarrow \infty$.
First, consider the limit of
\begin{equation}
 Z_k^\delta(s)
 = \sum_{n,n'} (n+1) (-1)^{\delta n} S_{nn'} \chi_k^{n'}(i s / \pi).
\end{equation}
By using Eq.(\ref{limit-character})
 with a condition of $s > s_0 \ne 0$
\begin{eqnarray}
 \lim_{k \rightarrow \infty} Z_k^\delta(s) &=&
 \lim_{k \rightarrow \infty}
 \sum_{n,n'} (n+1) (-1)^{\delta n}
 \cdot
 \sqrt{{2 \over {k+2}}}
 \sin \left( \pi {{(n+1)(n'+1)} \over {k+2}} \right)
\nonumber\\
 && \times
 {1 \over {(\eta(i s / \pi))^3}} 
 (n'+1)
 e^{- 2 (k+2) s \left({{n'+1} \over {2(k+2)}} \right)^2}.
\end{eqnarray}
The summation over $n$ and $n'$ can be carried out
 in the same way in the previous section, and we obtain
\begin{equation}
 \lim_{k \rightarrow \infty} Z_k^\delta(s) =
 \delta_{\delta 0}
 {1 \over {\left( 8 \pi^2 \alpha' \right)^{3/2}
                 \left( \eta( is / \pi) \right)^3}}
 \lim_{k \rightarrow \infty}
 2 \pi^2 \left( {2 \over Q} \right)^3,
\end{equation}
 where $Q=\sqrt{2/(k+2)}$ and
 we use $\alpha'=1/2$ in the open string Virasoro character
 due to the relation between a mode operator and 
 the space-time momentum operator:
 $\alpha_0^\mu =  \sqrt{2 \alpha'} p^\mu$.
Therefore,
 the flat space-time limit of the amplitude is
\begin{eqnarray}
 \lim_{k \rightarrow \infty} A^{M^6 \otimes W_k^{(4)}}
 &=&
 {1 \over 2}
 \cdot
 {1 \over {V_3}} 
 \lim_{k \rightarrow \infty}
 2 \pi^2 \left( {2 \over Q} \right)^3
\nonumber\\
 && \times
 {{i V_4} \over { 2 \pi \left( 8 \pi^2 \alpha' \right)^5}}
 \int_{s_0}^\infty ds
 \left( 8 \pi^2 \alpha' {\pi \over s} \right)^3
 e^{- {{y^2} \over {2 \pi \alpha'}} {\pi \over s}}
\nonumber\\
 &&\times
 \left( \eta(is/\pi) \right)^{-8}
 \left\{
  Z_\psi^{NS-NS}(i s / \pi) + Z_\psi^{R-R}(i s / \pi)
 \right\},
\label{amplitude-flat-limit}
\end{eqnarray}
 where $s_0$ is a cut-off by which
 the contribution from the higher modes of the closed string
 are eliminated.
The first factor $1/2$ is the remnant of the additional GSO projection
 which realizes the supersymmetry breaking
 by the fivebrane background.
The second factor simply reduces to unity,
 since $2\pi^2(2/Q)^3$ is the volume of the SU$(2)_k$ manifold
 of radius $2/Q \rightarrow \infty$
 in the $k \rightarrow \infty$ limit.
Therefore,
 the D3-brane tension
 which is obtained by taking the low-energy limit $s \rightarrow \infty$
 in the integrant of Eq.(\ref{amplitude-flat-limit})
 deviates from the one in the flat space-time (Eq.(\ref{tension-flat}))
 by a factor of $1/2$:
\begin{equation}
 \tau_3^2 \vert_{\rm flat-limit}
 = {1 \over 2} \cdot {\pi \over {\kappa^2}}.
\end{equation}
This deviation can be understood as a topological effect
 which may not disappear in the simple flat space-time limit.
Following the discussion in the first section,
 it is natural to consider that this deviation of the tension
 suggests the gravitino and/or dilatino condensation
 by the topological effect.

Here, we give a comment.
If $s$ is finite,
 we can take the flat space-time limit
 without using the formula of Eq.(\ref{modular-s}).
Namely,
 we can take the limit in the form of
\begin{equation}
 Z_k^{\delta}(\pi/s)
 = \sum_{n=0}^k (n+1) (-1)^{\delta n} \chi_k^n(i \pi / s).
\end{equation}
In this case we find that the amplitude,
 which includes a cut-off $s < s_\infty$,
 also reduces to the one in the flat space-time
 with an additional factor $1/2$.
Therefore, nothing special happens in the region of small $s$.
But,
 in order to take the flat space-time limit in the way
 which is consistent with the low-energy limit $s \rightarrow \infty$,
 we have to use the formula of Eq.(\ref{modular-s}).
 
\section{Conclusion}
\label{sec:conclusion}

The amplitude of the one closed string exchange
 between two parallel D3-branes in the fivebrane background
 is calculated.
The low-energy limit of the amplitude
 can be used to obtain the effective D3-brane tension
 in the low energy supergravity theory.
It has been shown that
 the tension from the flat space-time limit
 of the amplitude in the fivebrane background
 does not coincide with the one in the flat space-time.
Since the finite curvature effect
 should disappear in the simple flat space-time limit,
 the deviation suggests some topological effects.

The effective D3-brane tension in the flat space-time limit
 is the $1/2$ of the one in the flat space-time.
The origin of this factor $1/2$ is the additional GSO projection
 in the string theory in fivebrane backgrounds,
 as explained in section \ref{sec:fivebrane}.
This additional GSO projection
 results the spontaneous supersymmetry breaking
 in fivebrane backgrounds.
Therefore,
 it is natural to consider that
 the observed deviation of the effective D-brane tension
 is related with the supersymmetry breaking.

In the low energy supergravity theory
 the effective D-brane tension can be changed
 by the vacuum expectation value of some fields.
In case of the curved space-time background
 the finiteness of the curvature modifies the tension.
If the gravitino and/or dilatino condensation by the topological effect
 occurs as it is expected in Euclidean quantum gravity theory,
 it also modifies the tension.
Since the later modification due to the topological effect
 should not disappear in the simple flat space-time limit,
 we can understand that
 the observed deviation of the effective D-brane tension
 suggests the gravitino and/or dilatino condensation.
This conclusion is supported by the facts that
 the deviation is related with the supersymmetry breaking
 and the gravitino condensation triggers supersymmetry breaking.

In this analysis
 based on the intuitive picture of D-branes
 it is not clear whether the gravitino condensation really occurs
 and breaks supersymmetry.
The analysis
 in the language of the boundary states
 in the world-sheet conformal field theory\cite{RS,MMS,MMS2}
 must be important for further investigations.
The explicit form of the higher-order terms
 in the supersymmetric D-brane effective action must be also important
 to extract the actual values of condensates.

\acknowledgments

The author would like to thank S.~Saito and N.~Okada
 for fruitful discussions and comments.

\end{document}